\documentclass{Interspeech2024}

\usepackage{comment}
\usepackage{xcolor}
\usepackage{booktabs}
\usepackage{xcolor}
\usepackage{colortbl}
\usepackage{subcaption}
\usepackage{graphicx}
\usepackage{arydshln}
\usepackage{multirow}
\usepackage{amsmath}
\usepackage{amssymb}% http://ctan.org/pkg/amssymb
\usepackage{pifont}% http://ctan.org/pkg/pifont

\usepackage{hyperref}

\definecolor{almond}{rgb}{0.94,0.87,0.8} % almond
\definecolor{melon}{RGB}{232,156,124} % red
\definecolor{champagne}{rgb}{0.97, 0.91, 0.81}
\definecolor{baselinecolor}{rgb}{1,1,1}
\definecolor{corn}{rgb}{0.98, 0.93, 0.36}
\definecolor{teagreen}{rgb}{0.82, 0.94, 0.75}
\definecolor{predcolor}{rgb}{0.74, 0.83, 0.9}
\definecolor{decorrcolor}{rgb}{0.98, 0.85, 0.87} % palepink
\definecolor{camouflagegreen}{rgb}{0.47, 0.53, 0.42}
\definecolor{brightube}{rgb}{0.82, 0.62, 0.91} % violet
\definecolor{paletaupe}{rgb}{0.74, 0.6, 0.49}
\definecolor{pastelviolet}{rgb}{0.8, 0.7, 0.79}
\definecolor{knowcolor}{rgb}{0.80, 0.94, 0.75}
\definecolor{offlinecolor}{rgb}{1,1,1}
\definecolor{firsttaskcolor}{rgb}{0.97, 0.97, 0.97}
\definecolor{azure(colorwheel)}{rgb}{0.0, 0.5, 1.0}
\definecolor{gray(x11gray)}{rgb}{0.75, 0.75, 0.75}
\definecolor{lightgray}{rgb}{0.90, 0.90, 0.90}
\definecolor{darkgray}{rgb}{0.66, 0.66, 0.66}

\usepackage{graphicx}

\usepackage{tcolorbox}
\interspeechcameraready

\title{Efficient Fine-tuning of Audio Spectrogram Transformers via Soft Mixture of Adapters}

\name[affiliation={1}]{Umberto}{Cappellazzo}
\name[affiliation={2}]{Daniele}{Falavigna}
\name[affiliation={2}]{Alessio}{Brutti}

\address{
  $^1$University of Trento, Trento, Italy\\
  $^2$Fondazione Bruno Kessler, Trento, Italy }
\email{umberto.cappellazzo@unitn.it, \{falavi,brutti\}@fbk.eu} %third@companyB.ai}


\begin{document}

\maketitle

% the abstract here must exactly match the abstract entered into the paper submission system
\begin{abstract}

Mixture of Experts (MoE) architectures have recently started burgeoning due to their ability to scale model’s capacity while maintaining the computational cost affordable, leading to state-of-the-art results in numerous fields. While MoE has been mostly investigated for the pre-training stage, its use in parameter-efficient transfer learning (PETL) settings is underexplored. To narrow this gap, this paper attempts to demystify the use of MoE for PETL of Audio Spectrogram Transformers to audio and speech downstream tasks. Specifically, we propose Soft Mixture of Adapters (Soft-MoA). It exploits adapters as the experts and, leveraging the recent Soft MoE method, it relies on a soft assignment between the input tokens and experts to keep the computational time limited. Extensive experiments across 4 benchmarks demonstrate that Soft-MoA outperforms the single adapter method and performs on par with the dense MoA counterpart. We finally present ablation studies on key elements of Soft-MoA. Our code is available at \url{https://github.com/umbertocappellazzo/PETL_AST}.

\begin{comment}
Mixture of Experts (MoE) architectures have recently started burgeoning due to their ability to scale model's capacity while maintaining the computational cost affordable. Furthermore, they can be applied to both Transformers and State Space Models, the current state-of-the-art models in numerous fields. While MoE has been mostly investigated for the pre-training stage, its use in parameter-efficient transfer learning (PETL) settings is underexplored. To narrow this gap, this paper attempts to demystify the use of MoE for PETL of Audio Spectrogram Transformers to audio and speech downstream tasks. Specifically, we propose Soft Mixture of Adapters (Soft-MoA). It exploits adapters as the experts and, leveraging the recent Soft MoE method, it relies on a soft assignment between the input tokens and experts to keep the computational time limited. Extensive experiments across 4 benchmarks demonstrate that Soft-MoA outperforms the single adapter method and performs on par with the dense MoA counterpart. We finally present ablation studies on key elements of Soft-MoA, showing that Soft-MoA achieves better scaling with more experts, as well as ensuring that all experts contribute to the computation of the output tokens, thus dispensing with the expert imbalance issue.
\end{comment}

\end{abstract}

\section{Introduction}

Large pre-trained audio and speech models have exhibited outstanding performance when fine-tuned with task-specific data \cite{radford2023robust, gong2021ast}. A common practice entails the adaptation of the whole model to each downstream task (i.e., full fine-tuning) \cite{tsalera2021comparison, wang2021fine}. However, this paradigm has two major limitations: 1) adapting the entire pre-trained model is expensive and usually demands a significant volume of training data; 2) storing a copy of the model for each downstream task is unfeasible and impractical.

Given these shortcomings, current research mainly revolves around learning a small fraction of task-specific parameters, while keeping the pre-trained model frozen. This approach is known as \textit{parameter-efficient transfer learning} (PETL) and includes several nuances. For example, prompt-tuning methods \cite{jia2022visual, li2021prefix} introduce trainable task-specific tokens into one or multiple layers. LoRA \cite{hu2021lora, zeng2023expressive} uses trainable low-rank matrices to approximate the weight matrices. Adapter-based methods \cite{rebuffi2017learning, pfeiffer2020adapterfusion, jie2023revisiting} add lightweight modules (adapters) with bottleneck architecture comprising two fully-connected layers. The adapter can be inserted after both the multi-head self-attention and fully-connected feed-forward network blocks (\textit{Houlsby}) \cite{houlsby2019parameter}, or only after the feed-forward (\textit{Pfeiffer}) \cite{pfeiffer2020adapterfusion}. Adapters can also be scaled and shifted to modulate the pre-trained features \cite{lian2022scaling}, or their down/up projections can be shared across different layers and low-dimensional re-scaling coefficients are learned \cite{dong2023efficient}. In the speech field, PETL methods have been recently investigated and compared in \cite{cappellazzo2023parameter, lin2024peft}. 

Very recently, Mixture of Experts (MoE) models have shown remarkable results in natural language processing, pushing large language models to the limit, 
facilitating the effective scaling of Transformers and State Space Models while concurrently reducing computational costs \cite{mixtral, pushingMoE, deepseekmoe, moemamba}. The MoE paradigm relies on the idea that sub-modular components, the \textit{experts}, can specialize in different inputs and scale the model's capacity. While most works have focused on the use of MoE during the pre-training stage, only few works have leveraged MoE for efficient fine-tuning \cite{pushingMoE, adamix, adaptermix}. In the latter case, each expert is usually represented by a single adapter, and the model is referred to as Mixture of Adapters (MoA). However, these works usually target language-based tasks, whereas pure audio/speech classification tasks have not been taken into account before. Therefore, in this paper, we investigate the use of MoA for the Audio Spectrogram Transformer (AST), a powerful foundation model achieving state-of-the-art results on various audio/speech tasks \cite{gong2021ast}, and we ask the following question: 

\begin{tcolorbox}[colback=blue!10!white,colframe=blue!30!white,arc=2mm,outer arc=1mm]
%[colback=blue!5,colframe=blue!20,arc=2mm,outer arc=1mm]
\textbf{(}$\mathbf{Q}$\textbf{)} \textit{Can we leverage MoAs for the efficient fine-tuning of AST to audio/speech downstream tasks?}
\end{tcolorbox}

To answer the above research question \textbf{(}$\mathbf{Q}$\textbf{)}, we study the MoA's adoption for PETL of AST on four popular audio and speech benchmarks. Specifically, we propose to adapt a recent \textit{sparse} version of MoE called \textit{Soft-MoE} \cite{softMoE} to our PETL setting, whereby each expert only handles a small number of slots that are the result of a weighted combination of all input tokens. We call it \textbf{Soft-MoA}, and we compare it with the standard single adapter approach and with the dense version of MoA that requires each adapter to process all the input tokens (we refer to it as Dense-MoA). \textit{By doing this, we are able to scale the number of adapters while keeping the computational cost limited as well as updating only a small fraction of parameters, thus leveraging the strengths of both the MoE and PETL paradigms}. We empirically show that both Soft and Dense MoA outperform the single adapter approach, both for the Pfeiffer and Houlsby configuration, leading to accuracy improvement of up to 2.5\%; also, Soft-MoA attains performance parity with Dense-MoA while drastically trimming down the training cost. Finally, we further demonstrate the effectiveness of Soft-MoA by carrying out extensive ablation experiments revealing that \textbf{\ding{182})} both Soft and Dense-MoA gains over the single adapter strategy are more evident when fewer parameters are available, \textbf{\ding{183})} Soft-MoA is robust to ``\textit{expert imbalance}'', thus ensuring that all experts are involved in the learning process, and \textbf{\ding{184})} Soft-MoA attains the best performance accuracy when few slots ($1/2$) and several experts are used rather than the opposite case as multiple slots tend to learn redundant information.

\begin{figure*}[t]
    \centering
    \includegraphics[width=15cm]{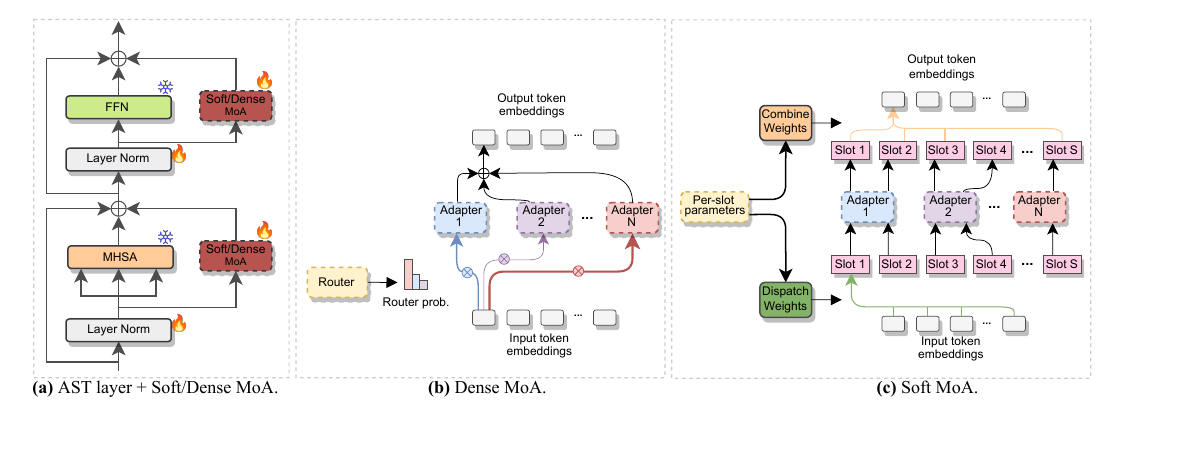}
    \caption{\textbf{(a)} For each AST layer, the Soft/Dense MoA blocks are inserted parallel to MHSA (Pfeiffer) or parallel to both MHSA and FFN sub-layers (Houlsby). \textbf{(b)} Illustration of Dense-MoA, whereby each expert contribution, scaled by the router's distribution (thickness of the arrows), is summed to produce the final output. \textbf{(c)} In Soft-MoA, each expert only processes a subset of slots (here $2$), and each slot accepts as input a weighted combination of all input tokens (thickness of the arrows). Note that the trainable parameters are represented by dashed blocks. Best viewed in color.}
    \label{fig:main_diagram}
\end{figure*}

\section{Methodology}
In this section, we first give a brief recap of the AST model and the standard single adapter approach. In section \ref{sec:DSMoA}, we present the details of the Dense-MoA and Soft-MoA approaches.

\subsection{AST model recap}
The Audio Spectrogram Transformer (AST) is an attention-based model that achieves state-of-the-art results on various audio and speech tasks \cite{gong2021ast, ssast}. The AST model receives as input audio spectrograms that are patchified and then a linear projection is applied to each patch. This results in a sequence of $L$ tokens of size $d = 768$, which we refer to as $\textbf{X} \in \mathbb{R}^{L \times d}$. AST comprises $12$ attention layers, each of which is composed of two sub-layers: a
multi-head self-attention (MHSA) and a fully-connected feed-forward (FFN) module. 

\subsection{Adapters}
Adapters are light subnetworks that are inserted into every layer of the AST model. To keep the parameters limited, adapters exploit a bottleneck architecture. The input sequence of hidden dimension $d$ is first down-projected into a low-dimensional space with size $r$ (the bottleneck dimension), and then up-projected back to the original dimension $d$. A non-linear activation function is also applied in-between the two fully-connected layers. 

While the bottleneck adapter is the most common design, recent works have also explored convolution-based adapters mainly for vision tasks (e.g., Convpass) \cite{convpass, convadapter}. In addition to this, adapters usually follow a Pfeiffer \cite{pfeiffer2020adapterfusion} or Houlsby \cite{houlsby2019parameter} configuration: the former places the adapter parallel or sequentially to the MHSA or FFN sub-layer, whereas the latter includes the adapter on both sub-layers.

\subsection{Dense and Soft MoA}
\label{sec:DSMoA}

\begin{table*}[t]
\centering
\caption{Performance evaluations of Dense and Soft-MoA on $4$ benchmarks for the Pfeiffer configuration. We report the top-1 accuracy for each dataset, the average over the four datasets (\textbf{Avg}), and the average train step time in milliseconds (\textbf{Time}).}
\label{tab:main}
\begin{tabular}{lccccccc}
\toprule
\textbf{Method} & \textbf{\# params} &\cellcolor{decorrcolor} \textbf{ESC-50} &\cellcolor{corn} \textbf{US8K} &\cellcolor{camouflagegreen} \textbf{GSC} &\cellcolor{melon} \textbf{FSC} & \cellcolor{predcolor}\textbf{Avg} & \cellcolor{champagne} \textbf{Time (ms)}\\
\midrule
\textcolor{darkgray}{Full FT} & \textcolor{darkgray}{85.5M} &\textcolor{darkgray}{87.48} &\textcolor{darkgray}{84.31} &\textcolor{darkgray}{97.31} &\textcolor{darkgray}{93.29} &\textcolor{darkgray}{90.07} & \textcolor{darkgray}{645} \\
Linear &9-40K &75.85 &77.93 &41.78 &27.52 &55.77 & 226 \\ 
\hline \addlinespace[2pt]
BitFit &102K &86.05 &82.17 &85.51 &63.85 &79.40 & 513 \\
%SPT &230K &84.30 &79.73 &75.28 &40.85 &70.04 & ? \\
DPT &230K &86.52 &83.67 &89.18 &68.60 &81.99 & 561 \\
Pref-T &221K &82.93 &81.39 &83.46 &55.75 &75.88 & 529 \\ 
LoRA &221K &86.45 &83.83 &93.61 &76.00 &84.97 & 525 \\
\hline
 \rowcolor{pastelviolet}
\multicolumn{8}{l}{\textbf{Bottleneck Adapter}}\\
\hline \addlinespace[2pt]

Single &470K &88.65 &83.36 &93.53 &78.19 &85.93 &513 \\
\textbf{D-MoA 14} &535K &89.55 &84.30&93.89 &82.43 &\textbf{87.54} & \textbf{\textcolor{red}{1689}} \\
\rowcolor{teagreen}
\textbf{S-MoA 14} &535K &89.08 &84.88 &93.91 &82.48 &\textbf{87.59}& 626  \\ 
\hline
\rowcolor{lightgray}
\multicolumn{8}{l}{\textbf{Convpass Adapter}}\\ 
\hline \addlinespace[2pt]
Single &491K &87.93 &83.38 &93.47 &77.62 &85.60 & 515 \\

\textbf{D-MoA 14} &535K &89.30 &84.32&93.70 &83.52 &\textbf{87.71} & \textbf{\textcolor{red}{1727}}  \\
\rowcolor{teagreen}
\textbf{S-MoA 14} &535K &88.43&84.29 &93.36 &80.36 &\textbf{86.61} & 638 \\ 

\bottomrule

\end{tabular}

\end{table*}

\textbf{Dense-MoA}. It encompasses a set of $N$ \textit{``expert'' adapters} $E_1, \dots E_N$ and a \textit{router network} $R$ that learns the optimal distribution over the adapters for a given input sequence. In its simplest form \cite{fedus2022switch, pushingMoE}, the router is a dense fully-connected layer with weights $\textbf{W} \in \mathbb{R}^{d \times N}$ followed by a \textit{softmax} function that takes as input the sequence $\textbf{X}$ and merges the output of each adapter using the gating scores $g_1, \dots g_N$ to yield the output sequence $\mathbf{Y}$:

\begin{equation}
    g_i = R(\mathbf{X})_i = \text{softmax}(\textbf{X}\textbf{W}),
\end{equation}
\begin{equation}
    \mathbf{Y} = \sum_{i=1}^{N}g_i \cdot E_i(\mathbf{X}).
\end{equation} 
If all the $N$ adapters take part in the computation of the output of a given input (scaled by the router's distribution), then we refer to this as \textit{Dense}-MoA (alternatively we can think of this as \textit{ensemble} MoA). Whereas this approach would cater to exact computation of gradients and end-to-end-learning, it would also incur a substantial increase in computational costs since each input token is computed by every expert rather than a single expert. To circumvent the above issue, we propose to adapt a recent method called \textit{Soft Mixture of Experts} \cite{softMoE} to our PETL setting where each expert is an adapter, and we call it \textit{Soft-MoA}. {Note that in our setting only the adapters are actually learned whilst the backbone model is frozen. 

\textbf{Soft-MoA}. Rather than feeding all input tokens to each expert, Soft-MoA passes a different weighted soft combinations of all input tokens to each expert. Unlike other sparse techniques like Top-$k$ \cite{topk} whereby only the $k$ experts that are assigned the highest router's probability are activated, Soft-MoA provides fully-differentiable operations, better training stability, and immunity to ``token dropping'' and ``expert imbalance'' issues \cite{softMoE}. In practice, each adapter processes $p$ slots, and each slot has a corresponding $d$-dimensional vector of parameters. These parameters are denoted by $\mathbf{\Phi} \in \mathbb{R}^{d \times (N\cdot p)}$. The input slots, $\mathbf{\tilde{X}}$, are computed as the convex combination of all the $L$ input tokens:  
\begin{equation}
    \mathbf{\tilde{X}} = \mathbf{D}^\top \mathbf{X}, \quad \mathbf{D}_{i,j} = \frac{\text{exp}((\mathbf{X}\mathbf{\Phi})_{i,j}}{\sum_{h=1}^{L} \text{exp}((\mathbf{X}\mathbf{\Phi})_{h,j})}.
\end{equation}
$\mathbf{D}$ is called the \textit{dispatch weights} and corresponds to applying a softmax along the columns of $\mathbf{X}\mathbf{\Phi}$. At this point, each adapter processes the corresponding slots: $\mathbf{\tilde{Y}}_i = E_{\lfloor i/p \rfloor}(\mathbf{\tilde{X}_i})$. Finally, the output tokens $\mathbf{Y}$ are the result of a convex combination of all $(N \cdot p)$ slots:
\begin{equation}
    \mathbf{Y} = \mathbf{C} \mathbf{\tilde{Y}}, \quad \mathbf{C}_{i,j} = \frac{\text{exp}((\mathbf{X}\mathbf{\Phi})_{i,j}}{\sum_{h=1}^{N \cdot p} \text{exp}((\mathbf{X}\mathbf{\Phi})_{i,h})}.
\end{equation}
The matrix $\mathbf{C}$ is referred to as the \textit{combine weights}, and is equivalent to applying a softmax over the rows of $\mathbf{X}\mathbf{\Phi}$. 

We provide an overview of Soft and Dense MoA in Figure \ref{fig:main_diagram}. %We show in Section \ref{sec:results} that Soft-MoA outperforms the single adapter approach and remains competitive with Dense-MoA while trimming down the computational cost. 
Finally, for our experiments, following \cite{cappellazzo2023parameter} that show that inserting the adapter in parallel achieves better performance than sequentially, we place the MoA block \textit{parallel} to the MHSA layer only (i.e., Pfeiffer) or \textit{parallel} to both the MHSA and FFN layers (i.e., Houlsby). The number of slots $p$ is an hyper-parameter, and we elaborate on its optimal value on Section \ref{sec:ablation}.

\section{Experiments and Discussion}

\subsection{Implementation Details}
For our experiments, we mainly follow the implementation details of \cite{cappellazzo2023parameter} to provide a fair comparison. 

\textbf{Datasets}. We evaluate the PETL methods on three audio/speech downstream classification tasks. (1) \textbf{Audio classification}: we use the ESC-50 and UrbanSound8K (US8K) datasets. ESC-50 \cite{piczak2015esc} consists of $2,000$ 5-second-long environmental audio recordings of $50$ classes. US8K \cite{salamon2014dataset} includes $8,732$ labeled sound excerpts of urban sounds from $10$ classes. (2) \textbf{Keyword spotting}: Speech Commands V2 \cite{warden2018speech} has $105,829$ 1-second recordings of $35$ speech commands. (3) \textbf{Intent classification}: Fluent Speech Commands (FSC) \cite{lugosch2019speech} includes $30,043$ English utterances spanning $31$ classes. 

\textbf{PETL baselines}. We include two traditional fine-tuning strategies: \textbf{full fine-tuning} (Full-FT), which finetunes the full pre-trained AST model; and \textbf{linear probing}, which only fine-tunes the classification head. Following \cite{cappellazzo2023parameter} we include some common PETL baselines: \textbf{BitFit} \cite{zaken2022bitfit}, \textbf{deep prompt-tuning} (DPT) \cite{jia2022visual}, \textbf{prefix-tuning} (Pref-T) \cite{li2021prefix} and \textbf{LoRA} \cite{hu2021lora}. For the analysis of MoA, we take into account both \textit{Bottleneck} \cite{houlsby2019parameter} and \textit{Convpass} \cite{jie2022convolutional} adapters. We report \textbf{Dense} and \textbf{Soft-MoA} (D/S-MoA) with $14$ or $7$ adapters for the \textit{Pfeiffer} and \textit{Houlsby} configuration, respectively, and we compare them with the standard implementation using a single adapter per layer (\textbf{Single}).

\textbf{Training Details}. For all experiments we use the AST model pre-trained on ImageNet-21K \cite{deng2009imagenet} and AudioSet \cite{audioset} provided by the Huggingface Transformers library \cite{huggingface}. The model has around $85.5$ million parameters, and the hidden size is $768$. Please refer to \cite{cappellazzo2023parameter} for the training details of the baselines (LoRA, DPT etc.). For MoA experiments, we use AdamW optimizer with cosine annealing scheduler and weight decay set to 0.1. For the ESC-50 and US8K datasets, we run 5-fold and 10-fold cross validation as suggested in the original papers. Except US8K that does not provide a validation set by default, for the others we set the hyper-parameters using the validation set. %Each experiment is carried out using a single A40/V100 GPU. The list of the hyper-parameters is available in our repository.

\subsection{Main Results and Discussion}
\label{sec:results}

\begin{table}[t]
\centering
\caption{Results of D/S-MoA for the Houlsby configuration. The number of parameters coincides with Pfeiffer as we still use 14 adapters split equally between MHSA and FFN layers.}
\label{tab:houlsby}
\begin{tabular}{lccccc}
\toprule
\textbf{Method} &\cellcolor{decorrcolor} \textbf{ESC-50} &\cellcolor{corn} \textbf{US8K} &\cellcolor{camouflagegreen} \textbf{GSC} &\cellcolor{melon} \textbf{FSC} & \cellcolor{predcolor}\textbf{Avg}\\
\midrule

\rowcolor{pastelviolet}
\multicolumn{6}{l}{\textbf{Bottleneck Adapter}}\\
\hline \addlinespace[2pt]
Single &88.00 &82.80 &91.75 &78.71 &85.32 \\
\textbf{D-MoA 7} &87.33 &83.78 &94.11 &82.64 &\textbf{86.97} \\
\rowcolor{teagreen}
\textbf{S-MoA 7} &87.13 &83.77 &93.67 &81.41 &\textbf{86.50}  \\ 
\hline
\rowcolor{lightgray}
\multicolumn{6}{l}{\textbf{Convpass Adapter}}\\ 
\hline \addlinespace[2pt]
Single &87.15 &82.75 &92.55 &77.79 &85.06 \\
\textbf{D-MoA 7} &87.31 &83.77 &93.20 &82.26 &\textbf{86.63}  \\
\rowcolor{teagreen}
\textbf{S-MoA 7} &88.13 &83.87 &92.69 &81.69 &\textbf{86.60} \\ 

\bottomrule

\end{tabular}

\end{table}

\begin{figure*}
\centering
\begin{subfigure}{0.335\textwidth}
    \includegraphics[width=\textwidth]{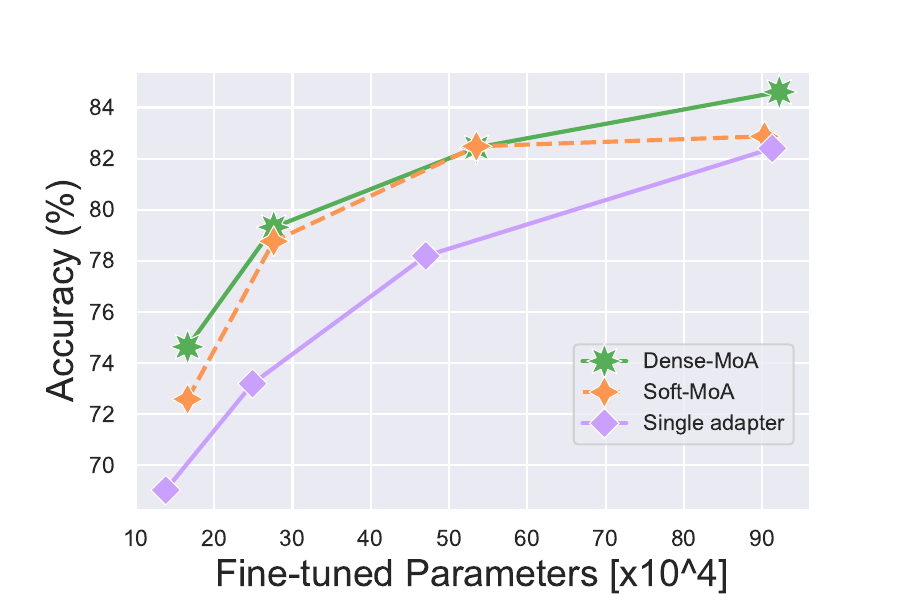}
\end{subfigure}
%\hfill
\begin{subfigure}{0.335\textwidth}
    \includegraphics[width=\textwidth]{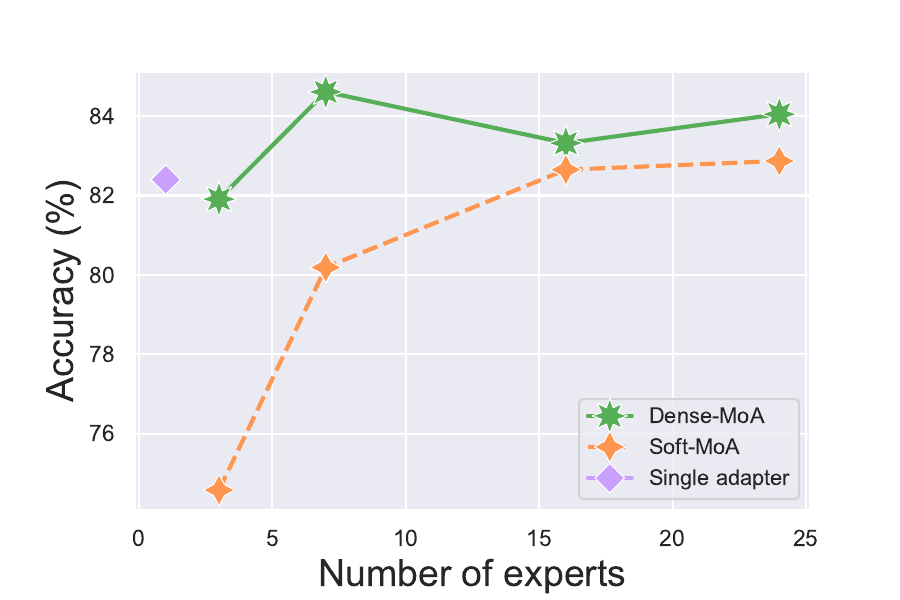}
    
\end{subfigure}
\begin{subfigure}{0.32\textwidth}
    \includegraphics[width=\textwidth]{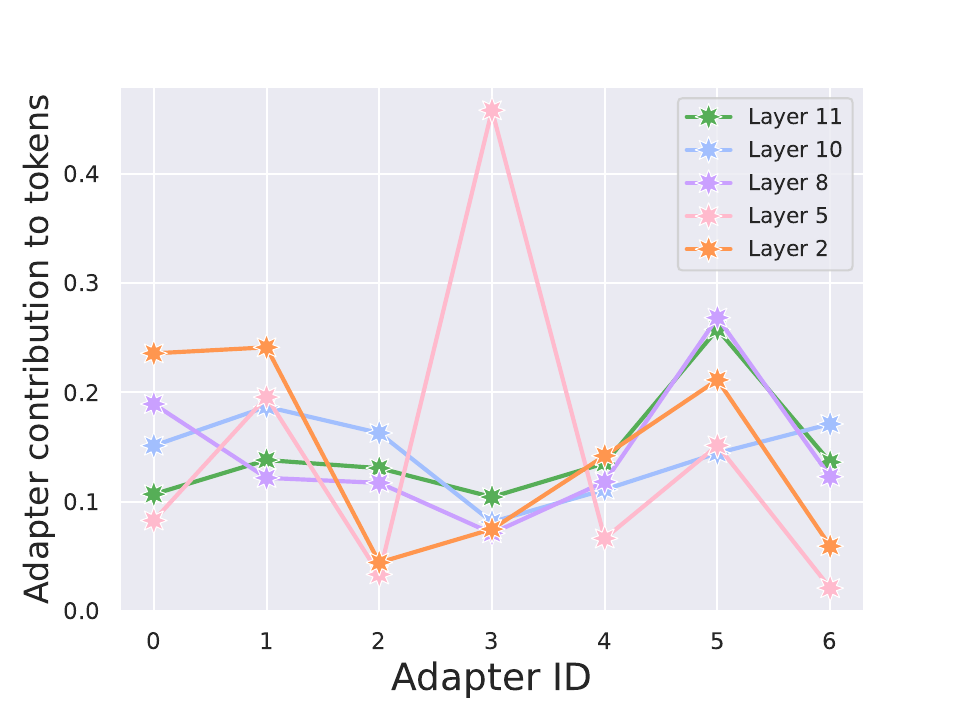}
\end{subfigure}
\caption{\textbf{(Left)}. The accuracy trend as more parameters are used. \textbf{(Middle)}. The effect of the number of adapters given a fixed parameters budget. \textbf{(Right)}. Adapters contribution to the output tokens for various layers. Results reported for FSC.}
\label{fig:vary_N}
\end{figure*}

Table \ref{tab:main} presents the performance comparisons between the single adapter approach and Soft/Dense MoA, as well as some other common PETL methods. The single adapter approach has bottleneck dimension equal to $24$, whereas Soft/Dense-MoA include 14 adapters, each with bottleneck dimension $1$, and one slot is used for each adapter. From table \ref{tab:main} we observe that both MoAs outperform the single adapter, leading to up to $2.5$ \% performance improvement on average for the Bottleneck case, while for Convpass we notice that Soft-MoA is slightly worse than Dense-MoA, but still better than the single adapter. In general, the biggest gain is obtained with the FSC dataset (up to $5.5$ and $7.6$ \%). Indeed, FSC is the more challenging dataset as it includes longer speech audio data, thus we argue that multiple adapters can specialize in learning different information, and consequently leading to better performance. We also notice that the GSC dataset does not benefit much from the use of MoA architecture. We surmise that a single adapter already achieves very competitive performance and so the use of multiple smaller adapters is not helpful. Another pivotal aspect is the extra computational cost brought by MoAs, estimated as the average train step time in milliseconds. Whereas Dense-MoA incurs a considerable increase in time (more than $3$x with respect to the single adapter), S-MoA, instead, requires only a limited extra time, while guaranteeing on-par performance. 

Finally, we test Soft-MoA's efficacy for the Houlsby configuration, where the MoA block is also inserted parallel to the FFN sub-layer. Table \ref{tab:houlsby} confirms the superiority of both MoAs over the single adapter.

\subsection{Ablation Studies}
\label{sec:ablation}
We now conduct some ablation studies to evaluate the effectiveness of Soft-MoA under different settings. We focus on the Pfeiffer Bottleneck configuration, and on the FSC dataset.

\textbf{Increasing the Parameters Budget.} We examine the methods' behaviour as we increase the number of trainable parameters. For the single adapter, we increase the parameters by making the bottleneck dimension $r$ larger, while for MoAs we keep it to $1$ and we increase the number of adapters. From Figure \ref{fig:vary_N} \textbf{(Left)} we observe that Soft-MoA outperforms the single adapter, although when more and more parameters are available the two methods tend to achieve similar results, thus showing that using a single adapter is a good alternative when scaling the number of parameters is sustainable.

\textbf{Few-big vs Many-small Adapters.} We now investigate how the MoA methods scale with respect to the number of adapters $N$. Regardless of $N$, we fix the number of learnable parameters to around $900$K to have a fair comparison. In this way, we want to figure out if having more adapters with a smaller bottleneck dimension is better than having a few but ``bigger'' (in terms of parameters) adapters. The Figure \ref{fig:vary_N} \textbf{(Middle)} shows that Dense-MoA, due to its intrinsic dense structure, reaches the peak performance when $N = 7$, and then adding more adapters does not lead to additional improvement. On the contrary, Soft-MoA depends heavily on $N$, and only when this number is large enough does it attain good performance. This trend is in line with that of the original Soft MoE paper \cite{softMoE}. 

\textbf{Adapters Contribution to the Output Tokens and Specific Classes.} By design, the computation of the final output tokens depends on a linear combination of all the adapters' slots. We want to verify whether all adapters contribute to the output sequence. We fix one slot per adapter and consider $7$ adapters, and we approximate the contribution of each adapter by averaging their coefficients in the linear combinations for all output tokens. We average over all the batches of the test set and report the adapter contribution for different layers in Figure \ref{fig:vary_N} \textbf{(Right)}. We see that some adapters have a bigger impact than others, but all of them contribute to the final output tokens. Therefore, Soft-MoA does not suffer from the expert imbalance issue, namely few adapters monopolize the output contribution while the others are overshadowed, an issue that affects other routing strategies like Top-$k$ \cite{topk, softMoE}. In addition to this, \textit{we compute the contribution of each adapter to each class}. To do this, for each sample of each class, we compute the contribution of each adapter and then we average over the total number of samples per class (for this reason the sum of each row of the heatmap does not sum to 1). We observe from Figure \ref{fig:heatmap} that some adapters specialize more for some classes than others (adapter $0$ has a high contribution for classes 26-29, adapter $1$ for classes $7$, $20$-$22$). We also see that the adapter with ID $5$ has a strong contribution for several classes.

\textbf{Optimal Trade-off between Slots and Adapters.} The number of slots $p$ is an important hyper-parameter of Soft-MoA, thus we examine its optimal value. We notice that if we set the number of slots equal to the number of tokens $L$, Soft-MoA boils down to Dense-MoA, so it is crucial to keep $p$ small. For our experiments, depending on the dataset, $L$ is between $100$ and $500$, and setting $p$ up to $14$ is a reasonable choice. We report the results for FSC in Table \ref{tab:ablation_adapterslot} and we see that, with the same number of trainable parameters, having more adapters with few slots brings better results than having few adapters but many slots. We speculate that this happens because multiple slots corresponding to the same adapter might have a tendency to learn similar concepts and become redundant, whereas using more adapters ends up learning more  diverse information.

\setlength{\tabcolsep}{10pt}
\begin{table}
\begin{minipage}[b]{0.40\linewidth}
\centering
\begin{tabular}{cc}
    \toprule
     \cellcolor{predcolor}\textbf{N/p}  &\cellcolor{predcolor} \textbf{Acc} \\
    \midrule
     2/14& 78.52 \\ 
    4/6 &80.26 \\ 
     6/4 & 81.65 \\
     8/3 & 82.36 \\
     \rowcolor{teagreen}
     \textbf{12/2} & \textbf{83.24} \\
     \rowcolor{teagreen}
     \textbf{24/1} & \textbf{82.87} \\
 \bottomrule
\end{tabular}
\caption{Optimal tradeoff between the number of adapters N and slots p.}
\label{tab:ablation_adapterslot}
\end{minipage}\hspace{1em}
\begin{minipage}[b]{0.55\linewidth}
\centering
\includegraphics[width=4.3cm]{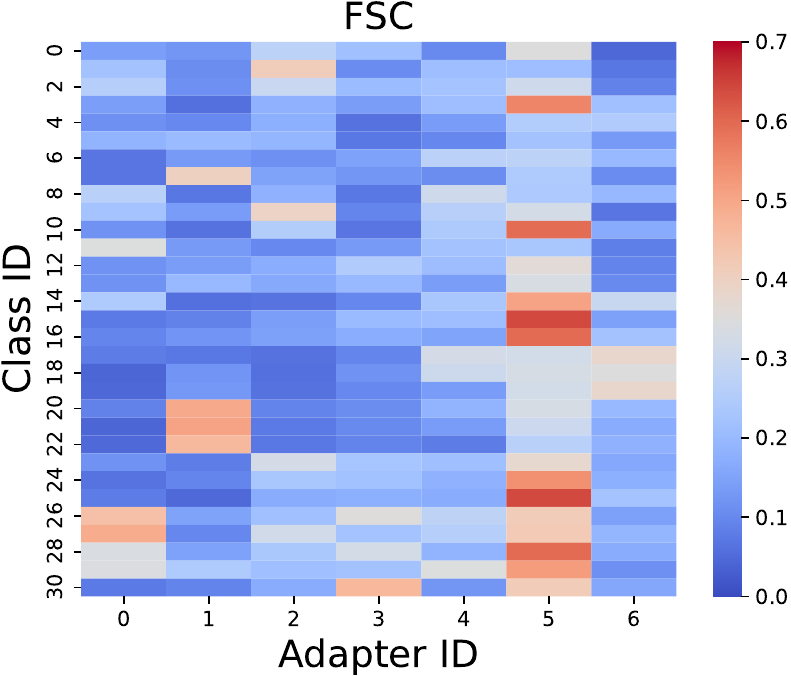}
\captionof{figure}{Distribution of expert activation frequencies per class.}
\label{fig:heatmap}
\end{minipage}
\end{table}

\section{Conclusion}

In this paper, we propose Soft Mixture of Adapters to efficiently fine-tune the AST model on various audio/speech downstream tasks. Soft-MoA relies on multiple adapters that take as input a soft convex combination of all the input tokens, thus reducing the computational cost of the dense counterpart. Extensive experiments on $4$ benchmarks show that Soft-MoA performs on par with Dense-MoA, and it outperforms the single adapter strategy, confirming itself as a strong method also for parameter-efficient transfer learning settings. To strengthen our analysis, we carry out ablation studies revealing that Soft and Dense MoA provide bigger gains over the single adapter when the parameters budget is limited. We also show that Soft-MoA scales better with the number of adapters and that it is sufficient to use only $1$ or $2$ slots to achieve the optimal performance.

\clearpage
\section{Acknowledgments}
Funded by the European Union. Views and opinions expressed are however those of the author(s) only and do not necessarily reflect those of the European Union or European Commission-EU. Neither the European Union nor the granting authority can be held responsible for them.

\bibliographystyle{IEEEtran}
\bibliography{mybib}

\end{document}